\documentclass[12pt]{article}
\usepackage{graphicx,amssymb,amsmath,times}

\title{An air shower array for LOFAR: LORA}
\author{S. Thoudam$^1$\thanks{s.thoudam@astro.ru.nl}, G. v. Aar $^1$, M. v.d. Akker$^1$, L. B\"ahren$^1$,\\
A. Corstanje$^1$, H. Falcke$^{1,2}$, J.R. H\"orandel$^1$, A. Horneffer$^1$,\\
C. James$^1$, M. Mevius$^3$, O. Scholten$^3$, K. Singh$^{1,3,a}$,\\
S. ter Veen$^1$\\
\small $^1$Department of Astrophysics, IMAPP, Radboud University Nijmegen, The Netherlands\\
\small $^2$ ASTRON, 7990 AA Dwingeloo, The Netherlands\\
\small $^3$ Kernfysisch Versneller Instituut, NL-9747 AA Groningen, The Netherlands\\
\small $^a$ Now at: IIHE, Vrije Universiteit Brussel, B-1050 Brussel, Belgium\\}

\begin{document}




\maketitle

\begin{abstract}
LOFAR is a new form of radio telescope which can detect radio emission from air showers induced by very high-energy cosmic rays. It can also look for radio emission from particle cascades on the Moon induced by ultra high-energy cosmic rays or neutrinos. To complement the radio detection, we are setting up a small particle detector array LORA (LOfar Radboud Air shower array) within an area of $\sim 300$ m diameter in the LOFAR core. It will help in triggering and confirming the radio detection of air showers with the LOFAR antennas. In this paper, we present a short overview about LORA and discuss its current status.
\end{abstract}


\section{Introduction}
LOFAR (the LOw Frequency ARray) is a new kind of radio telescope for astronomical observations in the frequency range of $\approx (10-240)$ MHz with high sensitivity and high spatial resolution (http://www.lofar.org). It uses a large number of simple dipole antennas instead of the traditional big parabolic dishes. It consists of $40$ stations in the Netherlands, $5$ in Germany and one each in Great Britain, France and Sweden covering a total area of more than $1000$ km in diameter.

Though primarily design as an astronomical telescope, LOFAR can also be used for the detection of very high-energy cosmic rays (CRs) in the interesting energy region above $10^{16}$ eV where the transition of galactic to extra-galactic CRs is expected (H\"{o}randel et al. 2009, Horneffer et al. 2010). This will be done by looking at extensive air showers which are essentially cascades of energetic secondary particles produce by the interaction of CR primaries with the nuclei present in the atmosphere. A large fraction of these secondaries are electrons and positrons which produce radio synchrotron emission in the presence of the Earth's magnetic field (Falcke et al. 2005). Due to coherence effects, this emission can give strong signals on the ground in the frequency range  of $\approx (10-80)$ MHz detectable by the LOFAR low band antennas. 

\begin{figure}[t]
\vspace*{2mm}
\begin{center}
\includegraphics[width=8.7cm]{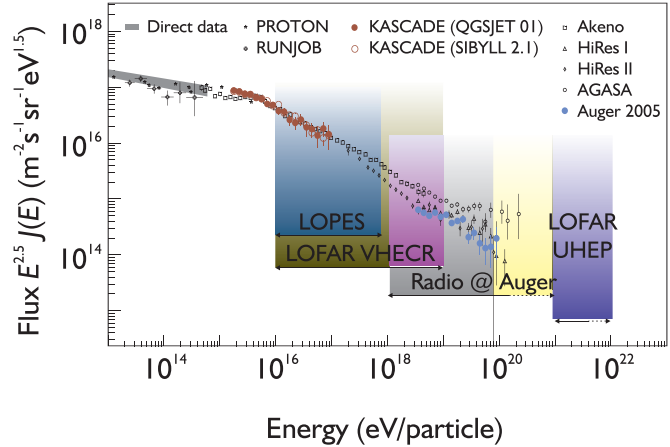}
\end{center}
\caption{Detectable CR energy ranges for LOFAR. LOFAR VHECR refers to the detection using air showers and UHEP to the detection using the Moon. Also shown are the energy ranges for the LOPES and the Auger radio experiments.}
\end{figure}

In addition, LOFAR can also look for the highest energy CRs or neutrinos above around $10^{21}$ eV (Scholten 2010). This will be carried out by detecting coherent radio Cherenkov emission from particle cascades at the Moon induced by those CRs or neutrinos. At frequencies of $\approx (100-200)$ MHz which is within the range of the LOFAR high band antennas, the angular spread of this emission becomes wider leading to more detectable signals at the Earth. The different energy ranges detectable by LOFAR are shown in Fig. 1 where the VHECR (Very High Energy CRs) refers to the detection using air showers and the UHEP (Ultra High Energy Particles) refers to that using the Moon. Fig. 1 also shows the energy ranges for the LOPES (LOFAR Prototype Station) experiment which is located at the KASCADE-Grande experimental site (Apel et al. 2010) and the currently building AERA (Auger Engineering Radio Array) experiment at the Pierre Auger Observatory (van den berg et al. 2009).

One important goal of the LOFAR CR experiment is to push the radio detection technique towards an independent way of detecting very high energy CRs. By detecting the radio signals with better sensitivity and better spatial resolution in a wider frequency range, we strongly believe that LOFAR will provide better understanding of the measured signals, their emission mechanisms and their relations with the air shower parameters, thereby leading to better estimates of the  properties of the primary particle. Compared to the LOPES (30 antennas) and the CODALEMA (24 antennas) experiments (Ardrouin et al. 2005), LOFAR has $18$ stations in its core (an area of $\approx 2\times 3$ km$^2$) with  each station consisting of $96$ low band antennas and $48$ high band antennas. However, at this stage, it is still quite early for radio detection experiments to do a stand-alone study on CRs. Therefore, we are also setting up a small particle detector array called LORA (LOfar Radboud Air shower array). Its main role will be to trigger and confirm the radio detection of air showers with the LOFAR antennas. It will also help in the reconstruction of several important air shower parameters like the primary energy, the shower core location, the arrival direction, the lateral density distribution etc. 

\section{LORA set-up}
LORA is setting up inside the LOFAR core within an area of $\sim 300$ m diameter. It consists of 5 stations with 4 particle detectors each, placed at a separation of $\sim (50-100)$ m between them. It is expected to detect CRs with energies greater than $\sim 10^{15}$ eV at an event rate of around once every few minutes. Detailed simulations about LORA as well as combined studies including the LOFAR radio data will be carried out as soon as possible. The schematic layout of the LORA  detectors arrangement is shown in Fig. 2 where the red stars represent the positions of the detectors and the circles denote the positions of the LOFAR antennas fields. Data collection in individual stations are controlled locally by station computers which are then controlled by a master computer where the overall data processing is done. 

\begin{figure}[t]
\vspace*{2mm}
\begin{center}
\includegraphics[width=7.0cm]{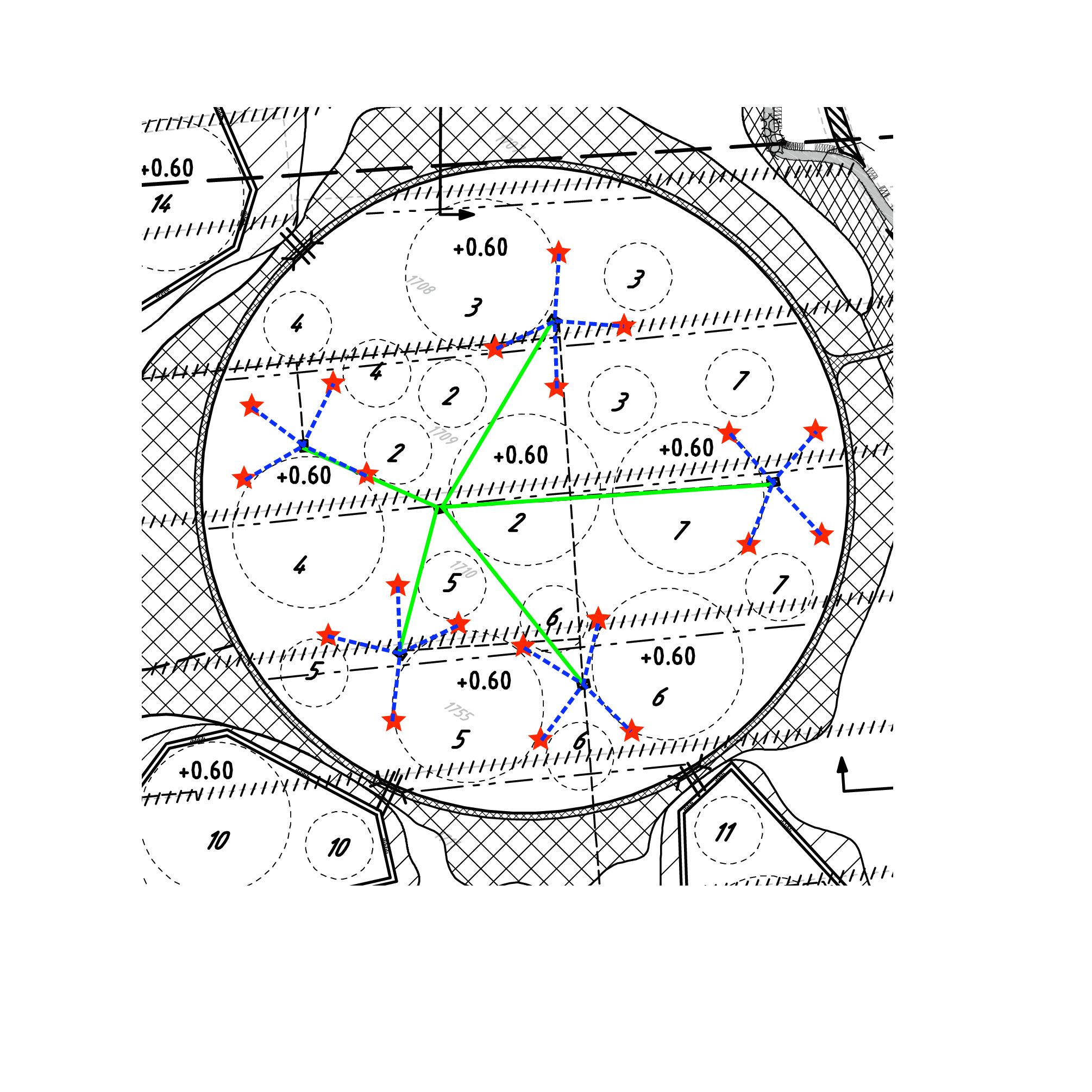}
\end{center}
\caption{Layout of the LOfar Radboud Air shower array (LORA) in the LOFAR core. The red stars denote the positions of the particle detectors and the circles denote the positions of the antenna fields. The detectors are contained within an area of $\sim 300$ m diameter with relative spacings of $\sim (50-100)$ m.}
\end{figure}

\subsection{Detectors}
The detectors for LORA are taken over from the KASCADE experiment (Antoni et al. 2003). Each detector is of $(98\times 125)$ cm in size and consists of two slabs of $3$ cm thick plastic scintillators. When a charged particle passes through a scintillator slab, the light which are produced are collected into a photomultiplier tube (PMT) through a wavelength shifter bar. The electrical signals from the PMTs (one PMT for each slab) are then combined and fed into the electronics where they are converted into digital signals. 

\subsection{Electronics}
The electronics was developed originally for the HISPARC experiment (http://www.hisparc.nl). An electronics unit (hereafter referred to as HISPARC unit) can handle two channels (one channel corresponds to one detector) and there are two units per station (a master and a slave). The master sets the event trigger condition and also provides the time stamp with a GPS receiver along with a $200$ MHz clock counter. Data in each channel are handled by two $12$-bit ADCs (so there are 4 ADCs per HISPARC  unit) which can measure voltages in the range of $(0-2)$ V. Event data (signal) in a channel are sampled with a time resolution of $2.5$ ns and are stored in a total time window of $10$ $\mu$s. Fig. 3 shows a typical signal of a charged particle passing through one of our detectors. Data from the electronics are sent to the station computer through USB.

Each HISPARC unit has a FPGA circuit built into it. This gives an observer control over several parameters required by the electronics as well as the detectors. For instance, one can set the trigger thresholds, high voltages, trigger condition, coincidence time window etc. at any time during the observation through a user software running at the station computer.

\begin{figure}[t]
\vspace*{2mm}
\begin{center}
\includegraphics[width=8.7cm]{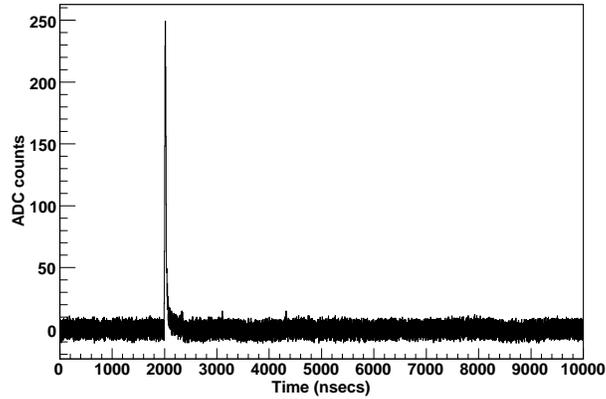}
\end{center}
\caption{Example of a signal trace when a charged particle passes through one of our detectors.}
\end{figure}

\subsection{Data acquisition}
All the PMTs are gain calibrated by adjusting their voltages to match the response for single muons. When the signals in the two HISPARC units in a station satisfy some minimum trigger condition, the signals are sent to the station computer which then sends them to the master computer. The master computer combines the signals from all stations and checks for good air shower events. When an air shower candidate is found, a first-level online analysis is performed to calculate shower parameters like the arrival direction, shower core position, event size etc., and a trigger information will be sent to the LOFAR antenna to dump the corresponding radio data for the air shower. The master computer takes less than $10$ ms to process an event. This is  necessary because the antennas have to dump their radio data corresponding to a particular event before they are overwritten in a memory ring-buffer, known as Transient Buffer Board (TBB). 

The DAQ software for LORA is developed in C/C++ language for Linux based operating systems. The software (particularly the online monitoring tool) uses several features of the ROOT package (http://root.cern.ch). On the monitoring panel, important information is displayed which is useful to monitor the performance of the system electronics and of the detectors during observations.

The final data are stored in ROOT format and they consist of four kinds of data. The first are the event data which are generated whenever an air shower event is detected. The data contain the event time stamp for each station and the signal trace in ADC counts for each detector. The second kind of data stores the so-called one second messages from the HISPARC unit. This data is generated every second by the master device and it contains information about the number of times the analog signal went over the threshold in the last second for each of the four channels. It also has important timing information which can be used for calculating an event time stamp with nanosecond accuracy. The third data kind comprises several control parameters for the observation run. This data is stored every interval of time fixed by the observer at the start of the run. The fourth kind of data contains information about the noise level in each channel averaged over some fixed interval of time.

The DAQ software and a preliminary data analysis software have been tested successfully on a LORA prototype installed at Radboud University (RU) Nijmegen. The results from the prototype are presented in the following section.

\begin{figure}[t]
\vspace*{2mm}
\begin{center}
\includegraphics[width=10.5cm]{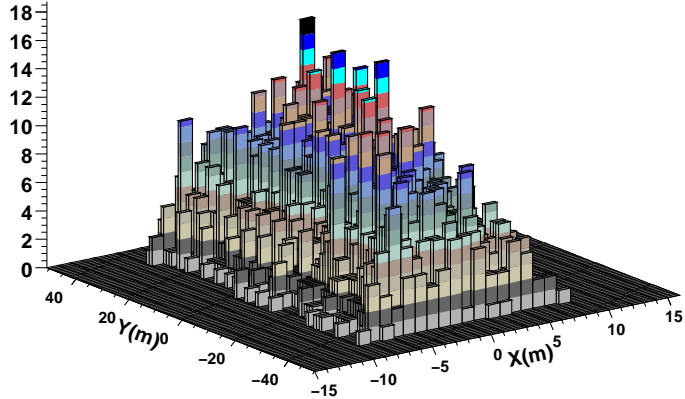}
\end{center}
\caption{Reconstructed air shower core distribution for 5463 events collected with the LORA prototype.}
\end{figure}

\begin{figure}[t]
\vspace*{2mm}
\begin{center}
\includegraphics[width=9.2cm]{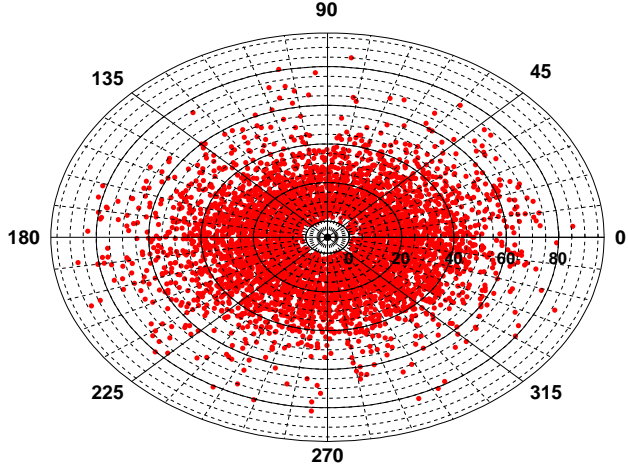}
\end{center}
\caption{Arrival direction distribution $(\theta,\phi)$ for the 5463 events collected with the LORA prototype.}
\end{figure}

\begin{figure}[t]
\vspace*{2mm}
\begin{center}
\includegraphics[width=8.7cm]{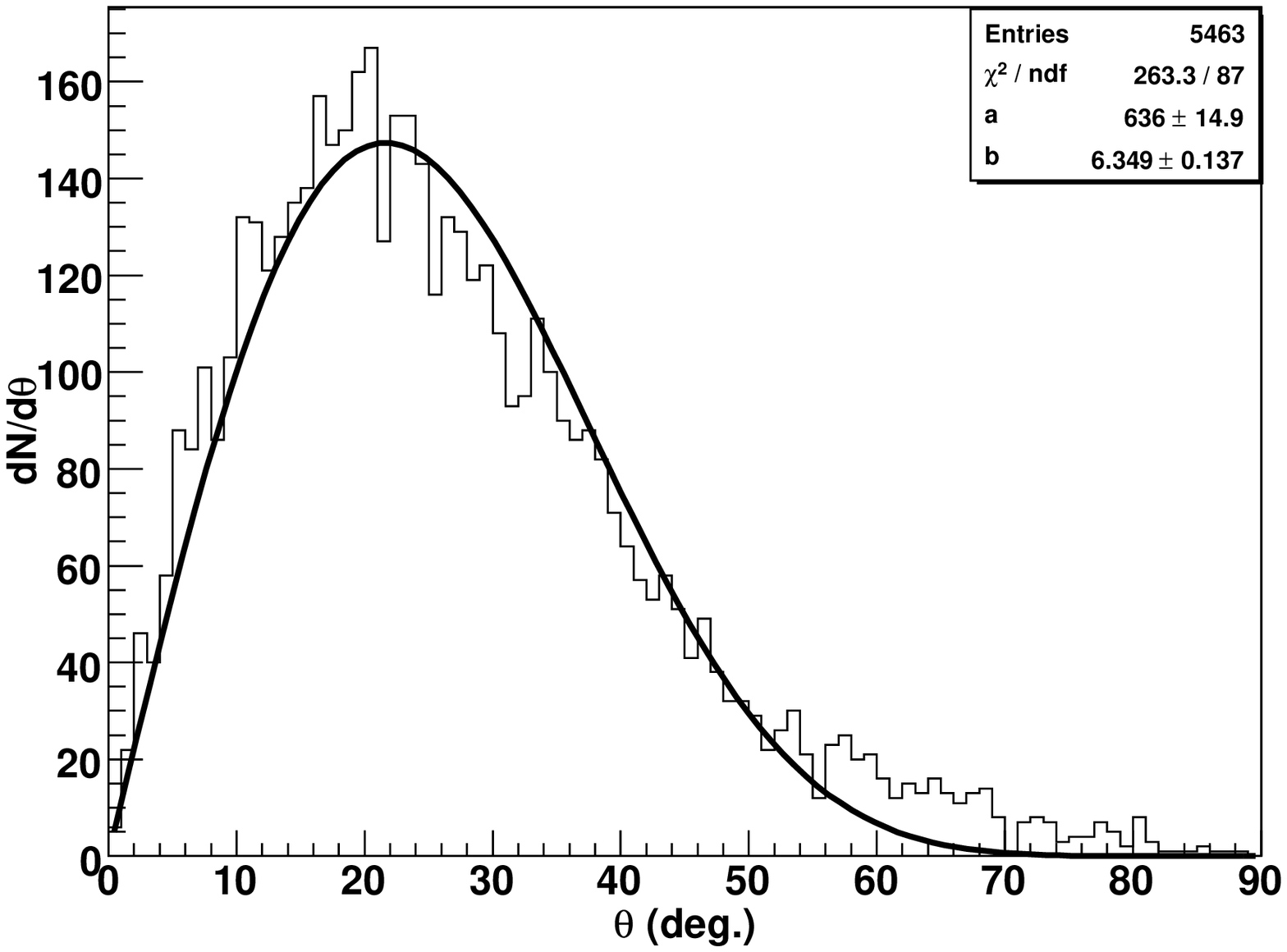}\\
\includegraphics[width=8.7cm]{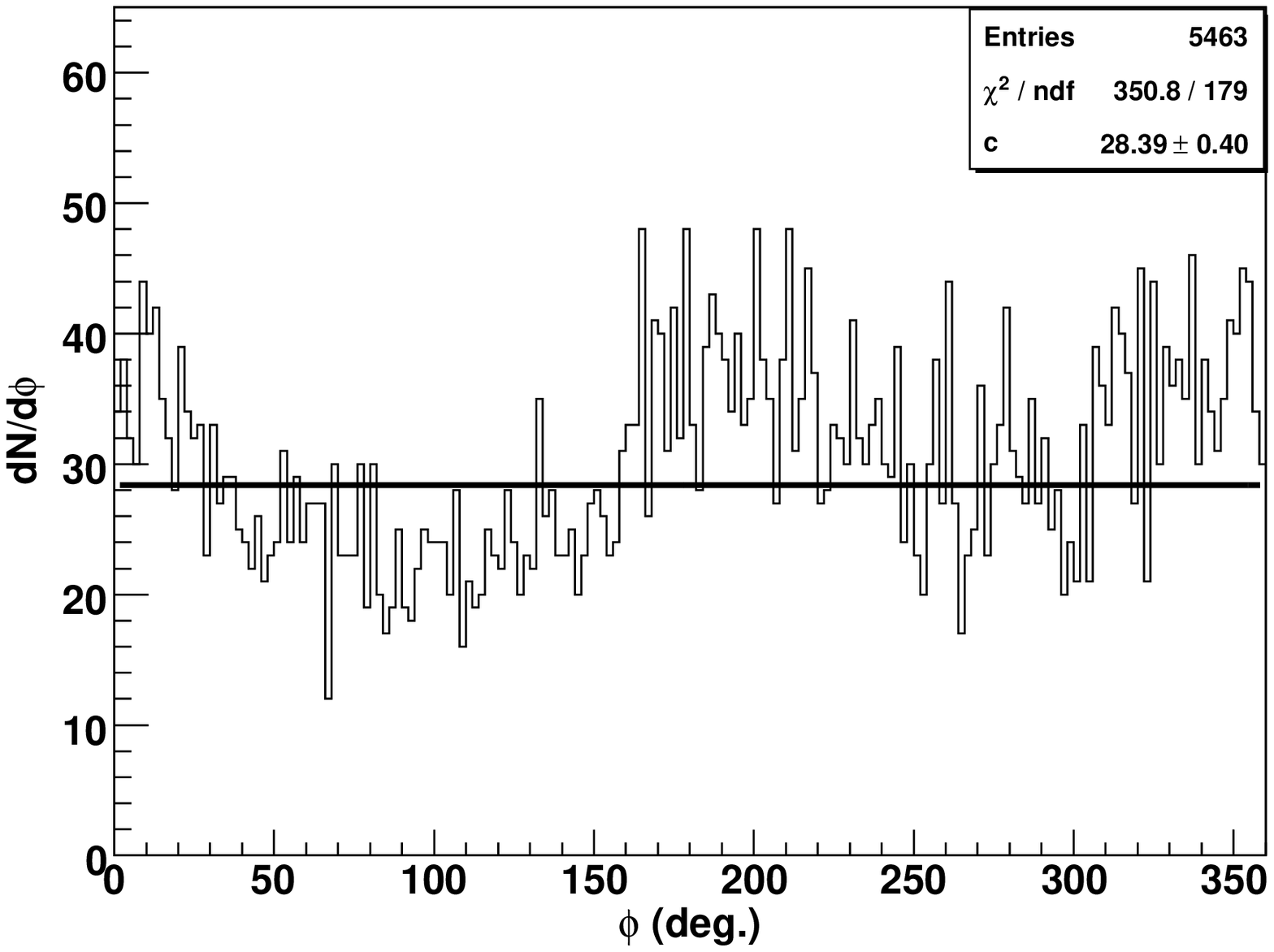}
\end{center}
\caption{Zenith $\theta$ (top) and azimuthal angles $\phi$ (bottom) distribution for the measured showers. The $\theta$ distribution is fitted with a function given by Eq. (1). The $\phi$ distribution is fitted with a constant.}
\end{figure}

\section{Results from the LORA prototype}
The LORA prototype that we set up to test the electronics and the DAQ software consists of 4 detectors in a $(20\times 80)$ m arrangement. With that, we collected more than $5000$ air shower events in a total observation time of $\sim 236$ hrs ($\sim 10$ days) during April $2010$. This corresponds to an event rate of around once every $3$ minutes. We have performed a preliminary analysis of the data to reconstruct the shower core position on the ground, the arrival direction and the energy of the primary. The primary energy is calculated from the total number of charge particles on the ground  using the parameterization given in H\"{o}randel 2007. This simple estimate gives the energy threshold of the prototype to be less than $10^{15}$ eV. The shower core position is calculated using the center of gravity method and the primary arrival direction is calculated using the relative pulse timing informations in the detectors assuming that the shower front arrives in a plane on the ground. The normal to the shower plane gives the arrival direction and it is characterized by two angles: the zenith angle $\theta$ measured from the vertical direction and the azimuthal angle $\phi$ measured in the horizontal plane. Though detailed simulation is still yet to be performed, the energy resolution of LORA is expected to be $\lesssim 30\%$ and the angular resolution to be $\lesssim 1^\circ$.

In Figs. 4 and 5, we show the distributions of the reconstructed core positions and the arrival directions $(\theta,\phi)$ respectively. In Fig. 6, we plot the $\theta$ (top) and the $\phi$ (bottom) distributions separately. The $\theta$ distribution is fitted using the following function,
\begin{equation}
\frac{dN}{d\theta}=a\;\mathrm{sin}\theta\;\mathrm{cos}^{b}\theta.
\end{equation}
The fit parameters are found to be $a=636\pm14.9$ and $b=6.35\pm 0.14$. For the $\phi$ distribution, we fit it to a constant as
\begin{equation}
\frac{dN}{d\phi}=c.
\end{equation}
Eq. (2) is expected if the arrival direction of the CRs are isotropic. However, in our case we see a sinusoidal distribution with two dips one at $\sim 90^\circ$ and the other at $\sim 270^\circ$ which are due to the narrow arrangement of our detector set-up.

It should be mentioned that at this stage, we do not aim to perform a detailed analysis and derive a final set of fit parameters which determine important properties of the CRs. As already mentioned, our primary aim was to test our electronics and the DAQ software. The preliminary results from the prototype are in general consistent with what we expect from CR observations with air shower arrays.

\section{Current status and future plans}
The electronics and the DAQ software for LORA have been tested successfully. The installation of all the LORA stations in the field is already completed and the set-up is currently under testing. On the other hand, the LOFAR CR pipeline software is under development and we are expecting to start the simultaneous observation of CRs with the radio antennas before summer 2011.\\
\\
\textbf{Acknowledgements}: The authors are deeply thankful to the KASCADE collaboration for kindly providing us some of the scintillators previously used for their experiment. The authors are also grateful to both the anonymous referees for constructive comments.














\end{document}